%% file: hbb52.tex
\documentclass[aps,prl,twocolumn,showpacs,superscriptaddress,groupedaddress,floatfix]{revtex4}
\usepackage{graphicx} 
\usepackage{dcolumn}  
\usepackage{bm}       
\usepackage{amssymb}  
\newcommand{\invisible}[1]{}

\def\tanb  {\ensuremath{\mathrm{\tan \beta}}\xspace}

\RequirePackage{xspace}
\def\b     {\ensuremath{b}\xspace}
\def\bbar  {\ensuremath{\overline b}\xspace}

\newcommand{\gevcc}{\ensuremath{{\mathrm{\,Ge\kern -0.1em V\!/}c^2}}\xspace}
\newcommand{\gevc}{\ensuremath{{\mathrm{\,Ge\kern -0.1em V\!/}c}}\xspace}
\def\bc   {\begin{center}}
\def\ec   {\end{center}}
\newcommand{\jprlBase}       {Phys.\ Rev.\ Lett.\xspace}
\newcommand{\jprl}      [1]  {\jprlBase\ {\bf #1}}

\begin{document}

\hspace{5.2in} \mbox{Fermilab-Pub-10-446-E}

\title{Search for neutral Higgs bosons in the multi-$\mathbf{b}$-jet topology in $\mathbf{5.2~fb^{-1}}$ of $\mathbf{p\bar{p}}$ collisions at $\mathbf{\sqrt{s} = 1.96}$ TeV}
\input author_list.tex 
\date{November 8, 2010}

\begin{abstract}
Data recorded by the D0 experiment at the Fermilab Tevatron Collider are analyzed to search for neutral Higgs bosons produced in association with $b$ quarks. The search is performed in the three-$b$-quark channel using multijet-triggered events corresponding to an integrated luminosity of $5.2~\mathrm{ fb}^{-1}$. In the absence of any significant excess above background, limits are set on the cross section multiplied by the branching ratio in the Higgs boson mass range 90 to 300 GeV, extending the excluded regions in the parameter space of the minimal supersymmetric standard model.
\end{abstract}

\maketitle 
The two Higgs boson doublets in the minimal supersymmetric standard model (MSSM)~\cite{mssm} lead to five physical Higgs bosons: three neutral (collectively denoted as $\phi$): $h$, $H$, and $A$; and two charged: $H^{+}$ and $H^{-}$. Two parameters, conventionally chosen as the ratio of the two Higgs doublet vacuum expectation values, $\tan\beta$, and the mass of the pseudoscalar $A$, $M_A$, are sufficient to describe the MSSM Higgs sector at tree level. Though $\tan\beta$ is a free parameter in the MSSM, there are indications which suggest that it should be large. A value of $\tan\beta \approx 35$ naturally explains the top to bottom quark mass ratio~\cite{topbottom}, and high $\tan\beta$ values also provide a good explanation for the observed density of dark matter~\cite{darkmatter}.

The couplings of the Higgs bosons to fermions in the MSSM are proportional to the corresponding couplings in the standard model (SM). The proportionality factor depends on the type of the quark (up- or down-type) and on the type of the Higgs boson. At large $\tan\beta$, the two Higgs bosons $A$ and either $h$ or $H$ have approximately the same mass and a down-type quark coupling enhanced by $\tan\beta$ compared to the SM coupling, while the coupling to up-type quarks is suppressed. Here, the three neutral Higgs boson couplings to $b$ quarks follow the sum rule $g_{hb\bar{b}}^2+g_{Hb\bar{b}}^2+g_{Ab\bar{b}}^2\approx 2\times\tan^2\beta \times g_{H,SM}^2$, where $g_{H,SM}$ is the SM coupling. Therefore, in these cases the production of Higgs bosons associated with bottom quarks (down-type quarks with the highest mass) is enhanced by a factor of $2\times\tan^2\beta$ compared to SM production. Due to the $\tan\beta$ enhancement, the main decay for the three neutral Higgs bosons is $\phi\to \b\bbar$ with branching ratios near 90\% (the remainder being mostly $\phi\to \tau\tau$). Since a direct search for $\phi \to \b\bbar$ is difficult due to large multijet backgrounds, searches rely on the case where $\phi$ is produced in association with one $b$ quark. The final state with three $b$ quarks represents a powerful search channel, with the third $b$-jet providing additional suppression of the large multijet background at a hadron collider.
    
MSSM Higgs boson production has been studied at the CERN LEP $e^+e^-$~collider which excluded \linebreak $M_{h,A}<93$~GeV for all $\tan\beta$ values~\cite{cite:LEP_exclu}. The CDF~\cite{cite:CDF_exclu} and D0~\cite{cite:D0_exclu, hbbrun2a, tautau} collaborations have extended MSSM Higgs boson searches to higher masses for high $\tan\beta$ values. This Letter uses data collected during Run II at the Fermilab Tevatron collider by the D0 collaboration corresponding to an integrated luminosity of $5.2~\mathrm{ fb}^{-1}$~\cite{d0lumi}. The dataset is broken into two periods, corresponding to the period before ($1.0~\mathrm{ fb}^{-1}$) and after ($4.2~\mathrm{ fb}^{-1}$) the upgrade of the D0 silicon vertex detector and trigger system. The dataset is five times larger than that used in the previous publication~\cite{hbbrun2a}. The full dataset has been reanalyzed to incorporate recent improvements to analysis procedures, algorithms, and calibrations. Improved modeling of the background has resulted in reduced systematic uncertainties. In addition, the Higgs boson mass range under consideration has been extended to 300 GeV. The limits are calculated using a program~\cite{cite:collie}, which is an implementation of the modified frequentist limit setting procedure~\cite{cls}, and are based only on the shape, and not the normalisation, of the distribution of the final discriminating variable. 

The D0 detector is described in Ref.~\cite{run2det}. Dedicated triggers for the three trigger levels (L1, L2, L3) designed to select events with at least three jets are used in this analysis. The majority of the data were recorded with $b$-tagging requirements at the trigger level, either at L3 or at both L2 and L3. The trigger has an efficiency of approximately 60\% for  $\phi\b \to \b\bbar\b$ events with a Higgs boson mass of 150 GeV when measured relative to events with three or four reconstructed jets.

The midpoint cone algorithm~\cite{cone} with a radius ${\cal R}=\sqrt{(\Delta y)^2 +(\Delta\varphi)^2}=0.5$, where $y$ is the rapidity and $\varphi$ the azimuthal angle, is used to reconstruct jets from energy deposits in the calorimeters. Details of the jet reconstruction and energy scale determination are described in Ref.~\cite{cite:jetx}. In addition to passing a set of quality criteria, all jets are required to be matched to at least two tracks reconstructed in the central detector, pointing to the $p\bar{p}$ vertex and with hits in the silicon detector. The matching criterion is $\Delta {\cal R}($track, jet-axis$)=\sqrt{(\Delta\eta)^2 +(\Delta\varphi)^2}<0.5$, where $\eta$ is the pseudorapidity. Signal events are selected by requiring three or four jets with transverse momenta $p_T > 20$ GeV and $|\eta| < 2.5$. The distance in the coordinate along the beam axis of the position of the $p\bar{p}$ vertex ($z_{PV}$) is required to be within 35 cm of the center of the detector. This is well within the geometric acceptance of the silicon detector, as needed for efficient $b$-tagging. A neural network (NN) $b$-tagging algorithm~\cite{bid}, which considers lifetime based information involving the track impact parameters and secondary vertices, is used to identify $b$-quark jets. Each event must have at least three jets satisfying a $b$-tag NN requirement. The single jet $b$-tagging efficiency is approximately 50\% for a light-quark jet mistag rate of 0.8$\%$. Data events with two tagged jets are used together with simulated events with two and three tagged jets to model the background. Finally, the transverse momenta of the two $b$-tagged jets with the highest $p_T$ are required to be above $25$ GeV. The data are split into four independent channels based on jet multiplicity (three or four jets) and running period.

The leading order event generator {\sc pythia}~\cite{pythia} is used to generate samples of associated production of $\phi$ and a $b$ quark in the 5-flavor scheme~\cite{cite:MCFM}, $gb\rightarrow \phi b$. The cross section~\cite{footnote1}, experimental acceptance, and the kinematic distributions of the $b$-quark jet produced in association with the Higgs boson are corrected to next-to-leading order (NLO) using {\sc mcfm}~\cite{cite:MCFM}. Multijet background events from the $b\bar{b}j$, $b\bar{b}jj$, $c\bar{c}j$, $c\bar{c}jj$, $b\bar{b}c\bar{c}$, and $b\bar{b}b\bar{b}$ processes, where $j$ denotes a light parton ($u$, $d$, $s$ quark or $\rm gluon$), are generated with the {\sc alpgen}~\cite{alpgen} event generator. A matching algorithm is used to avoid double counting of final states~\cite{matching}. The small contribution from $t\bar{t}$ production to the background is also simulated with {\sc alpgen}. Other processes, such as $Zb\bar{b}$ and single top quark production, are negligible. The {\sc alpgen} samples are processed through {\sc pythia} for showering and hadronization. All samples are then processed through a {\sc geant}-based~\cite{geant} simulation of the D0 detector. The same reconstruction algorithms are used for the simulated samples as for the data. A parameterized trigger simulation, based on efficiencies measured in data, is used to model the effects of the trigger requirements. The $b$-tagging is modeled by weighting simulated events based on their tagging probability measured using data~\cite{bid}. The efficiency of the requirements on the triggers, $z_{PV}$, and the number of jets, range from $1.9\%$ to $26.4\%$ for Higgs boson masses between 90 and 300 GeV. After the three $b$-tag requirement the efficiency with respect to the total number of signal events ranges from $0.2\%$ to $1.4\%$ in the three-jet channel ($0.1\%$ to $0.9\%$ in the four-jet channel). Table~\ref{tab:cutflow} shows the number of events in data and the signal efficiency at different stages of the event selection. 

\begin{table}[h]
\begin{small}
\begin{center}
 \begin{tabular}{ l c c c}
\hline \hline
                     & Number of events & Fraction (\%)  &Signal eff. (\%)\\
                     & ($\times 10^{3}$)&                & $M_A = 200$ GeV \\
\hline

Events   &      517,288      & 100     & -  \\
Trigger  &    198,106  & 38    & 30  \\
$z_{PV}$ cut    &  195,587 & 38 & 25     \\
$3$/$4$
jets   &  96,318/21,898    & 19/4.2 & 12/3.7      \\
2 $b$-tag jets  &   710/230 & 0.14/0.044  & 5.4/1.8 \\
3 $b$-tag jets  &   15/11   & 0.0029/0.0021  & 1.2/0.61 \\
\hline
\hline
\end{tabular}
\caption{\small The number and fraction of events in data and signal efficiency for each selection requirement.
As the data are split into three- and four-jet sub-samples, these numbers are reported separately in the last three rows. }
\label{tab:cutflow}
\end{center}
\end{small}
\end{table}

Multijet processes contribute to the background and the theoretical uncertainty on their cross sections is very large. The background composition is therefore determined by fitting distributions of the transverse momenta of jets of simulated events to data. The fractional contribution, $\alpha_{i}$, of the $i$th multijet background process is calculated from equations linking the $b$-tag efficiency for the $i$th background, $\epsilon_{k}^{i}$, with the number of observed events, $N_k$, where $k$ indicates the number of $b$-tagged jets (0--3) in an event~\cite{footnote2}, and the total number of events, $N_{\rm tot}$:

\begin{equation}
 \begin{array}{r c l}
  \sum_{\displaystyle i} \alpha_{i}                      &=& 1 \vspace{2mm}\\
  \sum_{\displaystyle i} \alpha_{i} \times\ \epsilon_{k}^{i}& =& N_{k}/N_{\rm tot}.
  \end{array}
\label{eq:system}
\end{equation}
The double $b$-tagged sample is dominated by $b\bar{b}j$, while the triple $b$-tagged sample consists of a mixture of approximately $50\%$ $b\bar{b}b$, $30\%$ $b\bar{b}j$, $15 \%$ $b\bar{b}c+bc\bar{c}$ and a remaining fraction consisting of $c\bar{c}j, bjj, cjj$, and $jjj$.

For every event, the two jet pairs with the largest scalar summed transverse momenta are considered as possible Higgs boson candidates. To remove discrepancies between data and simulation originating from gluon splitting ($g\rightarrow b\bar{b}$), jet pairs which do not fulfill $\Delta {\cal R} > 1.0$ are rejected. 

Six variables for which the data distributions are well modeled by the simulation are used to separate the jet pair from a Higgs boson from the background: $\Delta\eta$ between the two jets in the pair, $\Delta\phi$ between the two jets in the pair, the angle between the leading jet in the pair and the total momentum of the pair, the momentum balance in the pair~\cite{footnote3}, the combined rapidity of the pair, and the event sphericity. Based on these kinematic variables a likelihood discriminant, $\mathcal{D}$, is calculated according to:

\begin{equation}
    \mathcal{D}(x_1,....,x_6) =\frac{\prod_{i=1}^{6}{ P^{\rm sig}_i (x_i)}}{\prod_{i=1}^{6}{ P^{\rm sig}_i (x_i)}+\prod_{i=1}^{6}{ P^{\rm bkg}_i (x_i)}},
\end{equation}

\noindent where $P^{\rm sig}_i$ ($P^{\rm bkg}_i$) refers to the signal (background) probability density function (pdf) for variable $x_i$, and $(x_1,...,x_6)$ is the set of measured kinematic variables. The pdfs are obtained from triple $b$-tagged signal and simulated background samples. Two likelihood discriminants, providing discrimination in the Higgs boson mass ranges $90-130$ GeV (low-mass) and $130-300$ GeV (high-mass), respectively, are built by combining simulated signal samples from the appropriate Higgs boson mass ranges. Signal samples of equal size, interspaced by 10 GeV in $M_A$, are hence added together within the low-mass and high-mass range, respectively. After evaluating the likelihood, only the jet pairing with the larger $\mathcal{D}$ is kept for each event in each mass range. To further remove background from the final analysis sample, events are only selected if $\mathcal{D} > 0.65$. The likelihood requirements are optimized considering the variation of the predicted limit in \tanb. The final discriminant used in the limit calculation is the distribution of the jet pair invariant mass, $M_{b\bar{b}}$, after the selection requirement of the likelihood appropriate to the mass of the hypothesized Higgs boson.

\begin{figure*}[ht]
   \bc
   \includegraphics[width=0.45\linewidth]{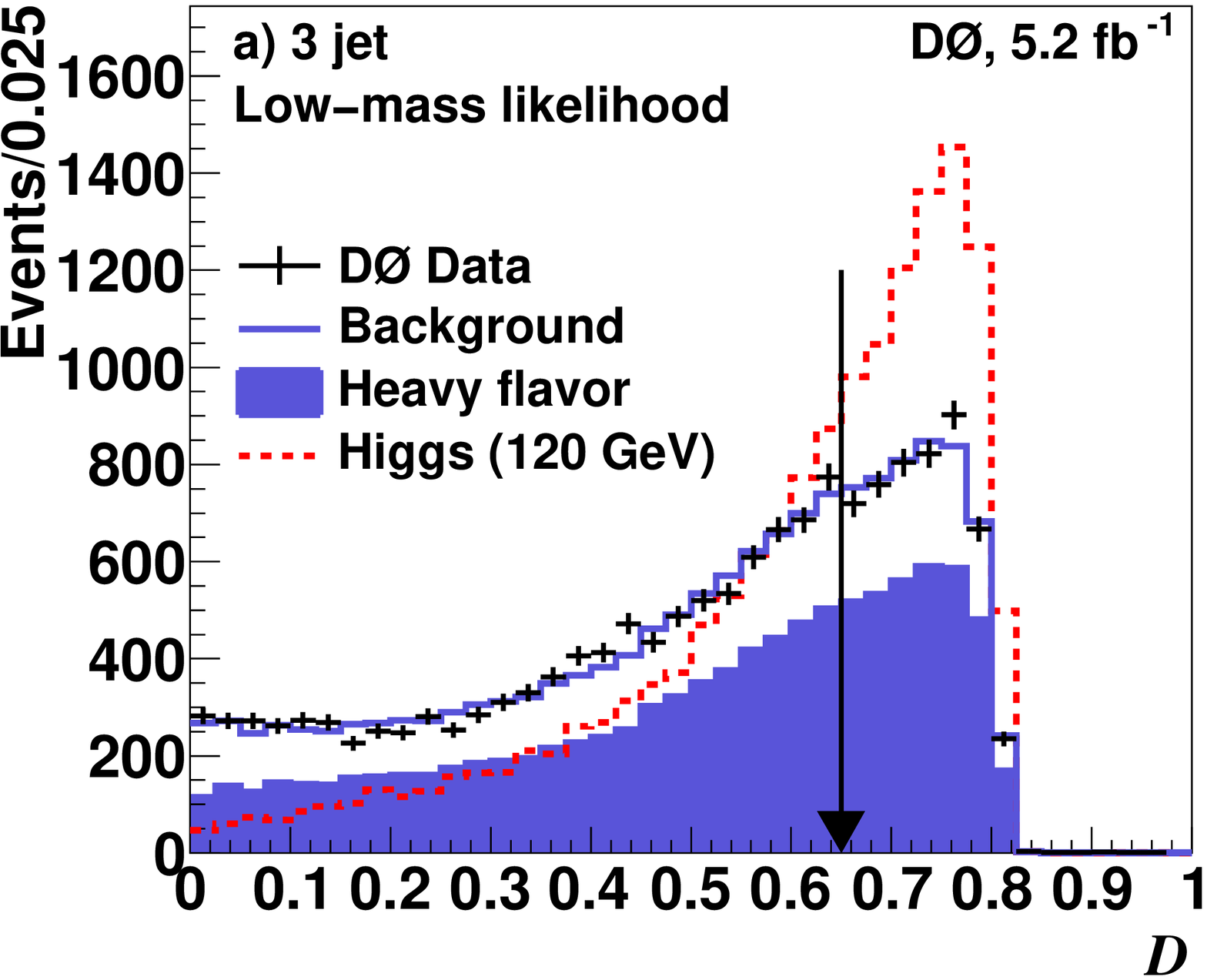}
   \includegraphics[width=0.45\linewidth]{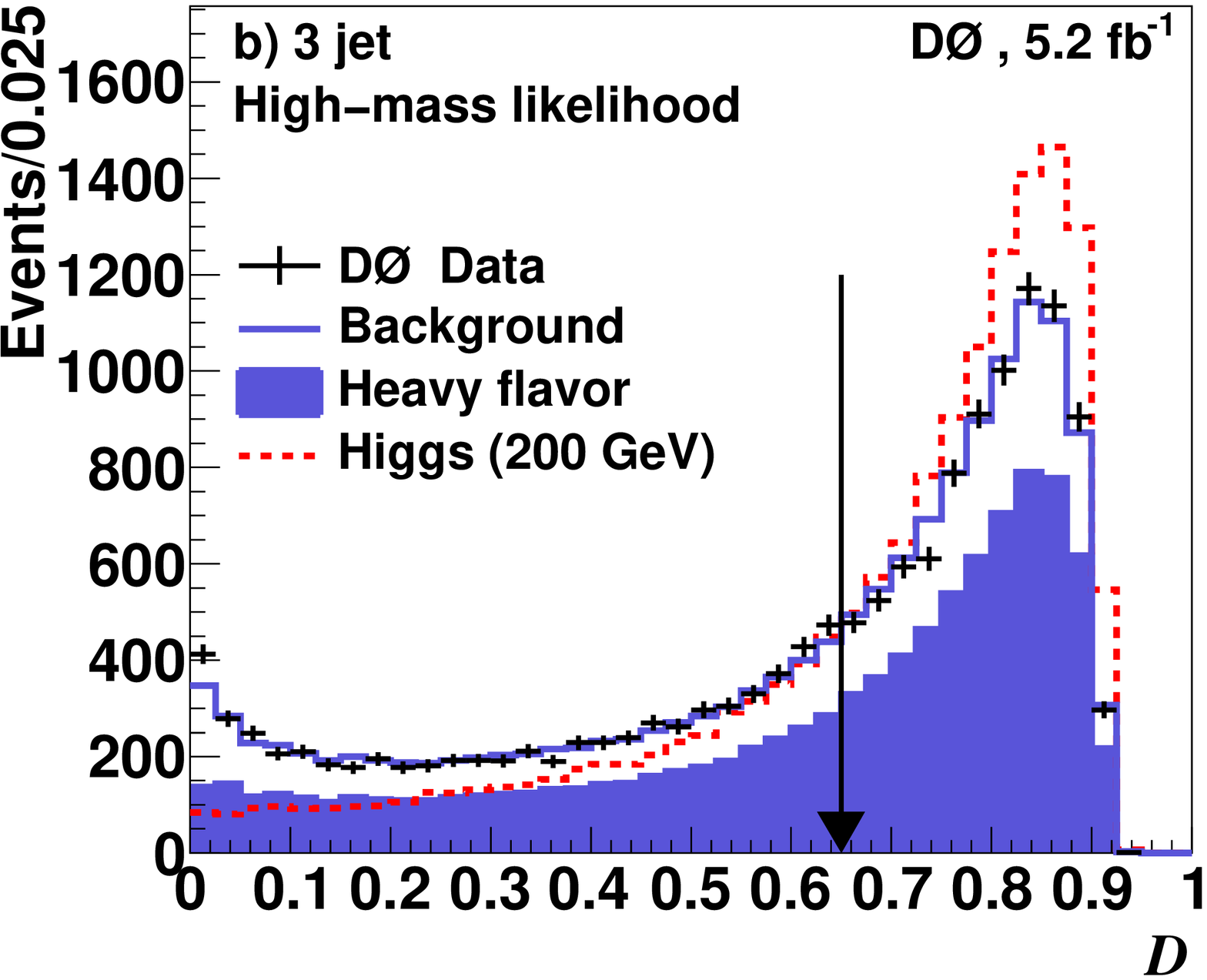}
   \caption{(color online) Comparison of the a) low-mass and b) high-mass likelihood distributions for the data and predicted background defined by Eq.~\ref{eq:BkgModel} in the 3Tag exclusive three-jet channel. Each event has one entry, the jet pairing with the highest likelihood output. Black crosses refer to data, the solid line shows the total background estimate, and the shaded region represents the heavy flavor component ($b\bar{b}b$, $b\bar{b}c$, and $bc\bar{c}$). The distributions for a Higgs boson of mass 120 and 200 GeV are shown as a dashed line in a) and b), respectively. The background and signal contributions are normalised to have an equal number of events as data. The arrows indicate the selection cut at $\mathcal{D} = 0.65$.}
   \label{fig:ThreeTagModel_YLH_MH100_p17}
   \ec
\end{figure*}

\begin{figure*}
   \bc
   \includegraphics[width=0.45\linewidth]{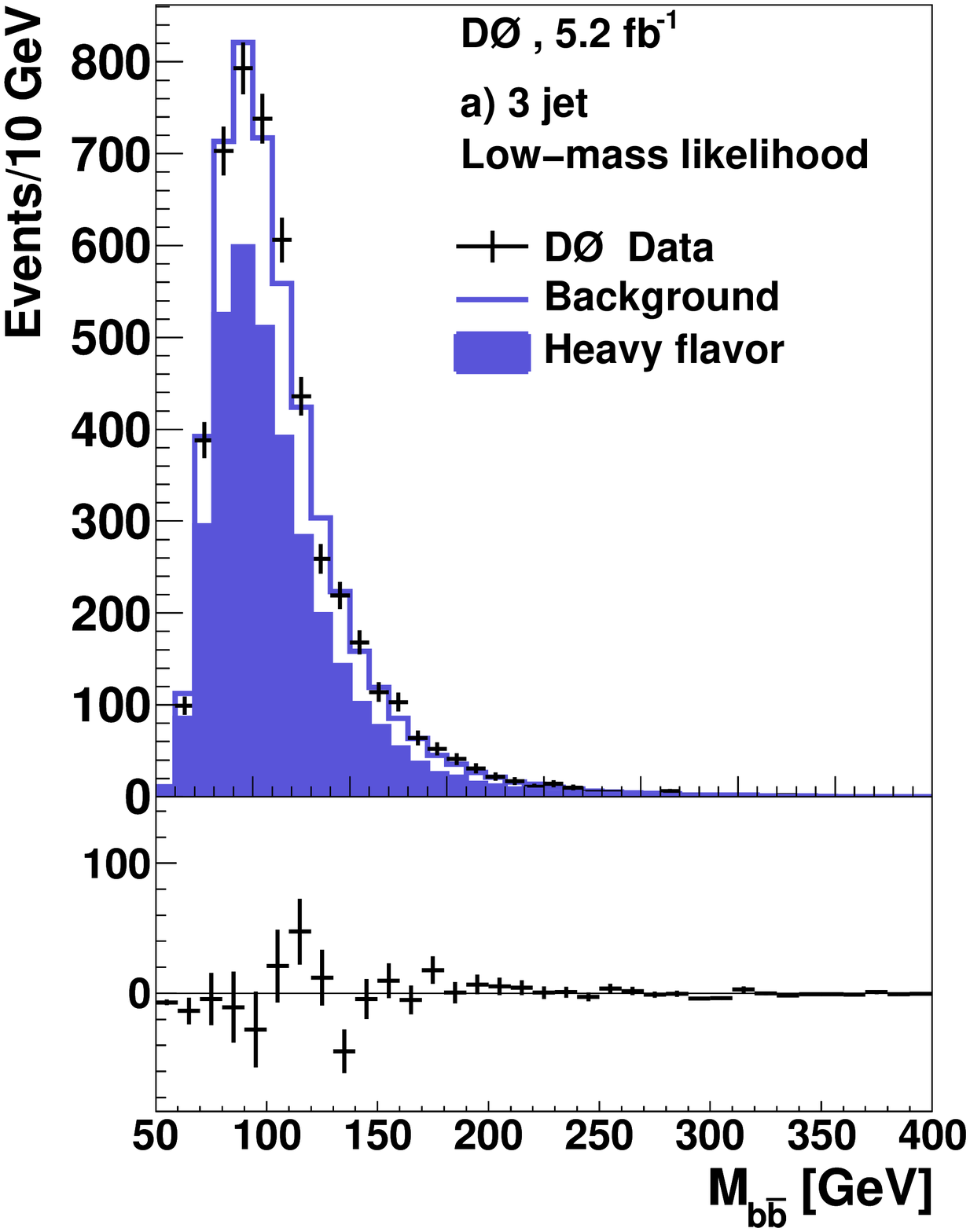}
   \includegraphics[width=0.45\linewidth]{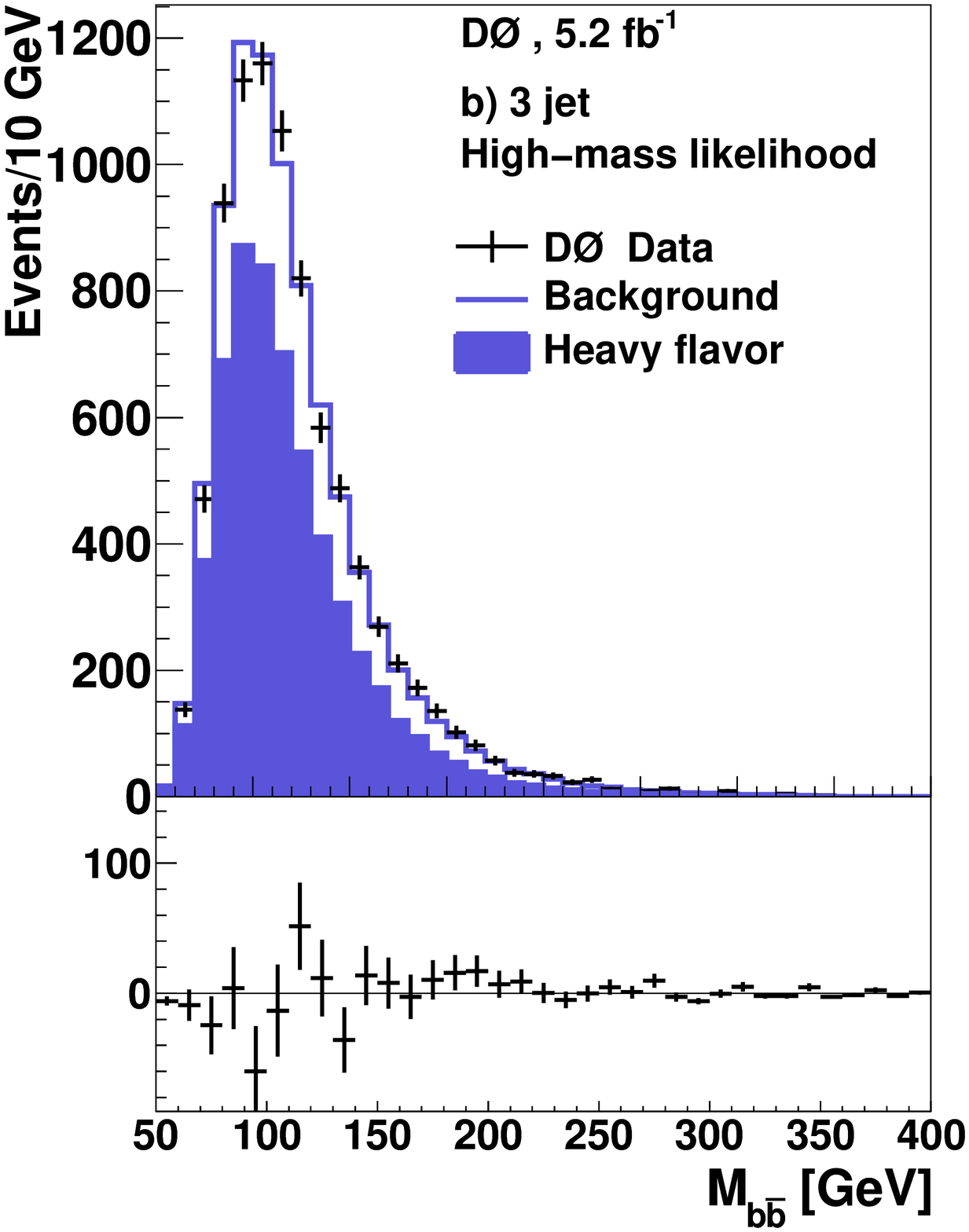}
   \caption{(color online) Invariant mass distribution for the 3Tag exclusive three-jet channel for: a) the low-mass likelihood selection and b) the high-mass likelihood selection. Each event has one entry, the jet pairing with the highest likelihood output. Black crosses refer to data, the solid line shows the total background estimate, and the shaded region represents the heavy flavor component ($b\bar{b}b$, $b\bar{b}c$, and $bc\bar{c}$). The lower panels show the difference between the data and the predicted background.}
   \label{fig:mass}
   \ec
\end{figure*}

\begin{figure}[ht]
   \bc
\includegraphics[width=0.9\linewidth]{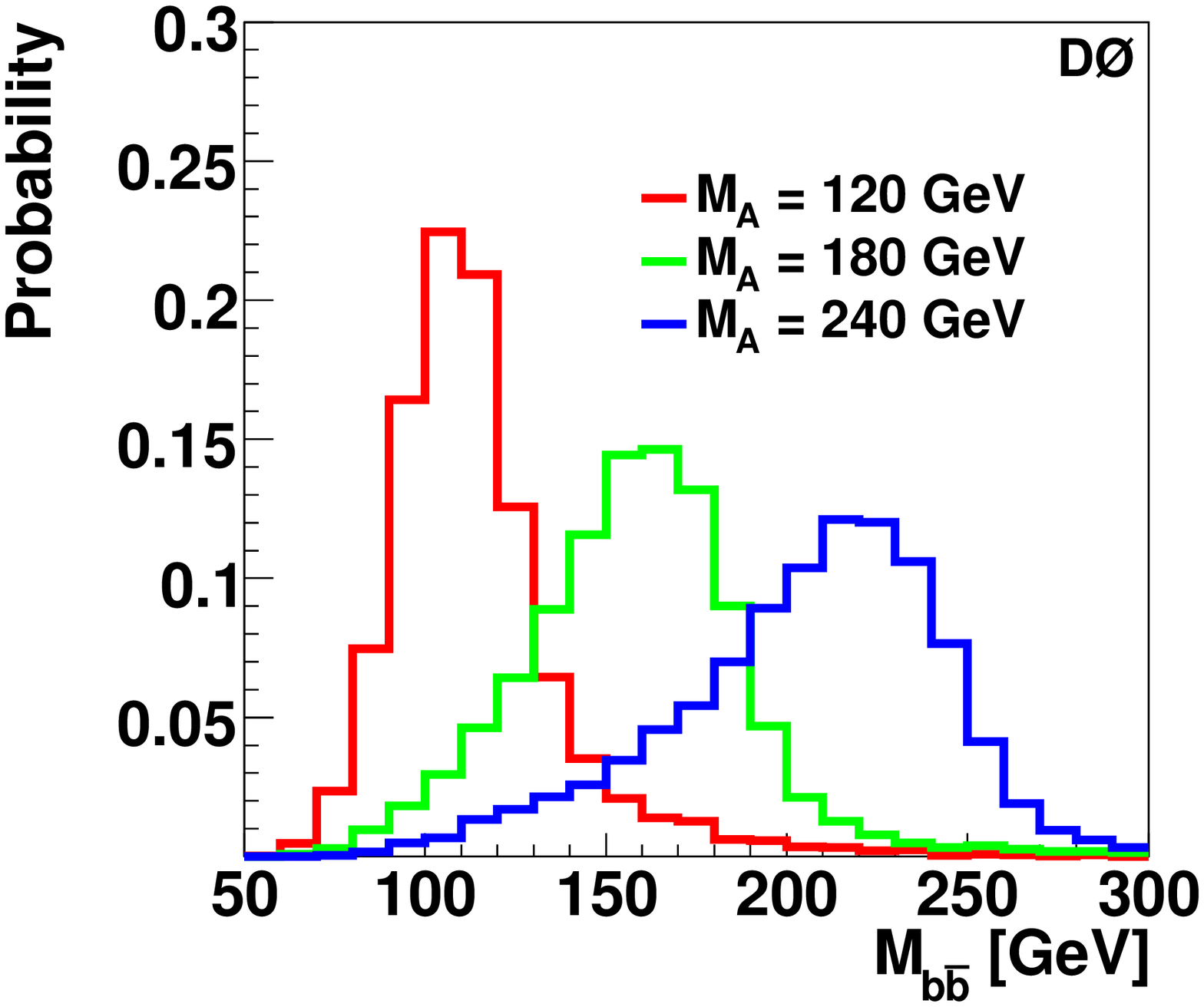}
   \caption{(color online) Invariant mass distribution of Higgs signals in the three-jet channel for $M_A =$ 120, 180 and 240 GeV, respectively. The distributions are normalised to unit area.}
   \label{fig:mass2}
  \ec
\end{figure}

The $b\bar{b}b$ background is indistinguishable from the signal events where the wrong $b$-jet pair is chosen by the likelihood and consequently cannot be normalised from the data. The $M_{b\bar{b}}$ background shape is modeled using a combination of data and simulated samples. The distribution, $S_{\rm 3Tag}^{\rm pred}(\mathcal{D},M_{b\bar{b}})$, of the predicted triple $b$-tagged (3Tag) background sample in the two-dimensional $\mathcal{D}$ and $M_{b\bar{b}}$ plane is obtained from the inclusive double $b$-tagged (2Tag) data shape multiplied by the ratio of the simulated shapes of the SM triple ($S_{\rm 3Tag}^{\rm MC}$) and double tagged events ($S_{\rm 2Tag}^{\rm MC}$):

\begin{equation}
S_{\rm 3Tag}^{\rm pred}(\mathcal{D},M_{b\bar{b}}) =\frac{S_{\rm 3Tag}^{\rm MC}(\mathcal{D},M_{b\bar{b}})}{S_{\rm 2Tag}^{\rm MC}(\mathcal{D},M_{b\bar{b}})}S_{\rm 2Tag}^{\rm data}(\mathcal{D},M_{b\bar{b}}).
   \label{eq:BkgModel}
\end{equation}
Fig.~\ref{fig:ThreeTagModel_YLH_MH100_p17} shows $\mathcal{D}$ for data and background for the low- and high-mass likelihoods in the three-jet channel. The shape of a signal distribution, normalised to the same number of events as data, is also shown. Fig.~\ref{fig:mass} shows the $M_{b\bar{b}}$ distribution in the three-jet channel after the low- and high-mass likelihood selection requirements, respectively. The invariant mass of a Higgs boson signal in the three-jet channel is shown in Fig.~\ref{fig:mass2} for three different values of $M_A$.

\begin{figure}[ht]
   \bc
\includegraphics[width=0.9\linewidth]{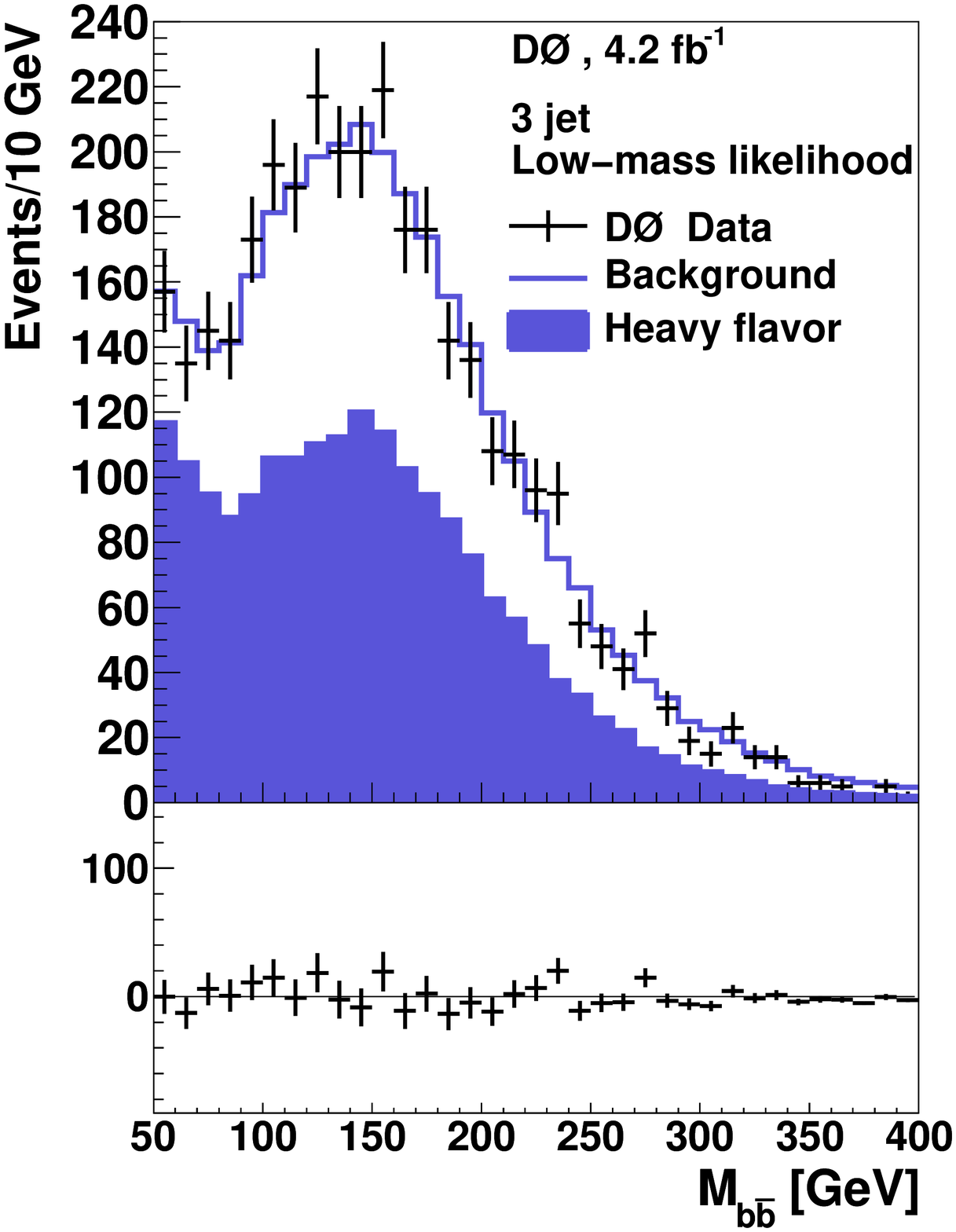}
   \caption{(color online) Invariant mass distribution for the jet pair with the lower likelihood for data (black crosses) and background (line) in the 3Tag exclusive three-jet channel for the 4.2~fb$^{-1}$ data set collected after the silicon detector upgrade. The lower panel shows the difference between the data and the predicted background.}
   \label{lowlh}
  \ec
\end{figure}

To verify the background model a signal-depleted region is studied - any deviation observed there is unlikely to be as a result of signal and therefore would indicate a possible problem in the background modeling. A sample is hence chosen using the lower likelihood jet pairing and applying a selection of $\mathcal{D} < 0.12$. Fig.~\ref{lowlh} shows the invariant mass distributions for background and data in this sample. Agreement ($\chi^2/ \mathrm{n.d.f.}$ = 0.86) between the background model and the data is observed. A wide variety of additional cross-checks were carried out, examining aspects of the event selection, b-tagging, and background modelling; no significant changes in the results were observed.

Sources of systematic uncertainty on both the signal normalisation and shape are considered. The sources of systematic uncertainty on the signal included are: $b$-quark jet identification efficiency,  $b$- and light-quark jet energy resolution, trigger modeling, jet energy calibration, jet identification, integrated luminosity, and theoretical models. The theoretical uncertainty on the signal cross section is estimated from {\sc mcfm}~\cite{cite:MCFM} and consists of a contribution of 10\% from the choice of factorisation scale as well as an uncertainty of $5\%$ to $13\%$ from the parton distribution functions, depending on Higgs boson mass. Both the theoretical uncertainty and the luminosity uncertainty of 6.1\%~\cite{d0lumi} are treated as normalisation uncertainties for each mass hypothesis. The remaining sources of systematic uncertainty are assessed individually by varying parameters within their uncertainties and taking into account the resulting difference in normalisation and shape of the $M_{b\bar{b}}$ distribution at each mass point.

For the dominant background, only systematic uncertainties affecting the shape of $M_{b\bar{b}}$ matter, since only the shape and not the normalisation is used to distinguish signal from background in this analysis. Additionally, many uncertainties affecting the simulation, like the jet energy scale and resolution uncertainties, cancel in Eq.\ref{eq:BkgModel}. The estimated variations in the remaining systematic sources are propagated to $\mathcal{D}$, $M_{b\bar{b}}$ and the predicted shape $S_{\rm 3Tag}^{\rm pred}$ and used in assessing the limits presented below. The uncertainty from the $b$-tagging of jets is evaluated by varying the $b$-tag efficiencies within their uncertainties. The uncertainties in the difference of the energy resolution between heavy and light flavor jets is obtained by smearing the energy of the $b$- and $c$-quark jets by an additional 7\%, corresponding to half the light-quark jet energy resolution. 
The shape difference between triple and double $b$-tagged data in the trigger turn-on curves resulting from the $b$-tagging criteria in the trigger is accounted for as a systematic uncertainty. Small variations in the shape, arising from possible signal contamination when determining the background composition, are included as a systematic uncertainty. Finally, the uncertainty on the $t\bar{t}$ normalisation is taken as 10\%~\cite{top}. 

No significant excess over the background is observed in the data. Limits on the Higgs boson production cross section multiplied by the branching ratio to $b\bar{b}$ are therefore calculated with the modified frequentist method~\cite{cite:collie,cls}. The confidence level of the signal, CL$_s$, which is used to calculate the exclusion, is defined in Ref.~\cite{cls}. The overall normalisation of the background expectation is allowed to float independently in the null (background-only) and test (background-plus-signal) hypotheses. The systematic uncertainties on the signal and background are included in the limit setting procedure. Each component of systematic uncertainty is adjusted by introducing multiplicative scale factors and maximizing the likelihood for the agreement between prediction and data with respect to these scale factors, constrained by prior Gaussian uncertainties. Limits on the product of cross section and branching ratio are obtained by scaling the signal cross section until $1-{\rm CL}_s = 0.95$ is reached. These limits are effectively independent of the signal model but assume the width of the $\phi$ to be negligible relative to the experimental resolution ($\approx$20\% at $M_A$ = 150 GeV). The four independent analysis channels are combined in the limit setting procedure. Signal hypotheses are considered for discrete Higgs boson mass points from 90 to 300 GeV in steps of 10 GeV. The treatment of the systematic uncertainties and the limit setting procedure were extensively cross-checked; no unexpected effects were observed.  

\begin{figure}[h]
   \begin{center}
\includegraphics[width=0.9\linewidth]{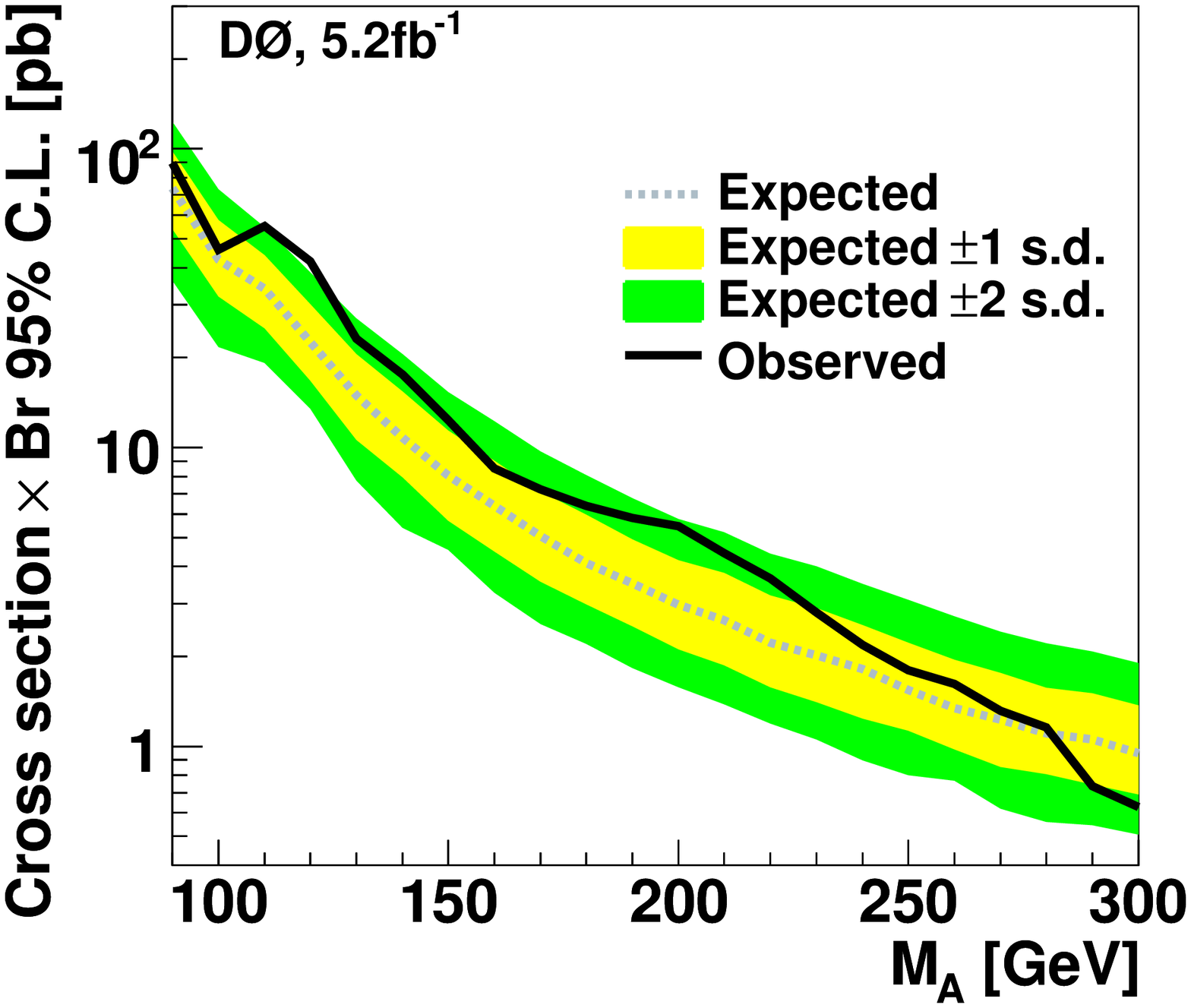}
   \caption{(color online) Model independent 95\% C.L. upper limit on the cross section multiplied by the branching ratio for the combined 5.2 fb$^{-1}$ analysis. The light and dark grey regions correspond to the one and two standard deviation (s.d.) bands around the expected limit.}
  \label{fig:p20Limit-mod-indep}
  \end{center}
\end{figure}

The combined result is summarized numerically in Table \ref{tab:xsecComb} and the model independent limit is shown in Fig.~\ref{fig:p20Limit-mod-indep}. The deviation from expectation around 120 GeV corresponds to 2.5 standard deviations. Note that it is more likely to find a deviation (in the background-only hypothesis) when several mass bins are probed than if only one bin is probed. A standard convention to account for this ``trial factor''~\cite{lookelse} gives a significance of the deviation at 120 GeV of 2.0 standard deviations.

\input xsec-table2.tex

\begin{figure*}[ht]
   \bc
\includegraphics[width=0.45\linewidth]{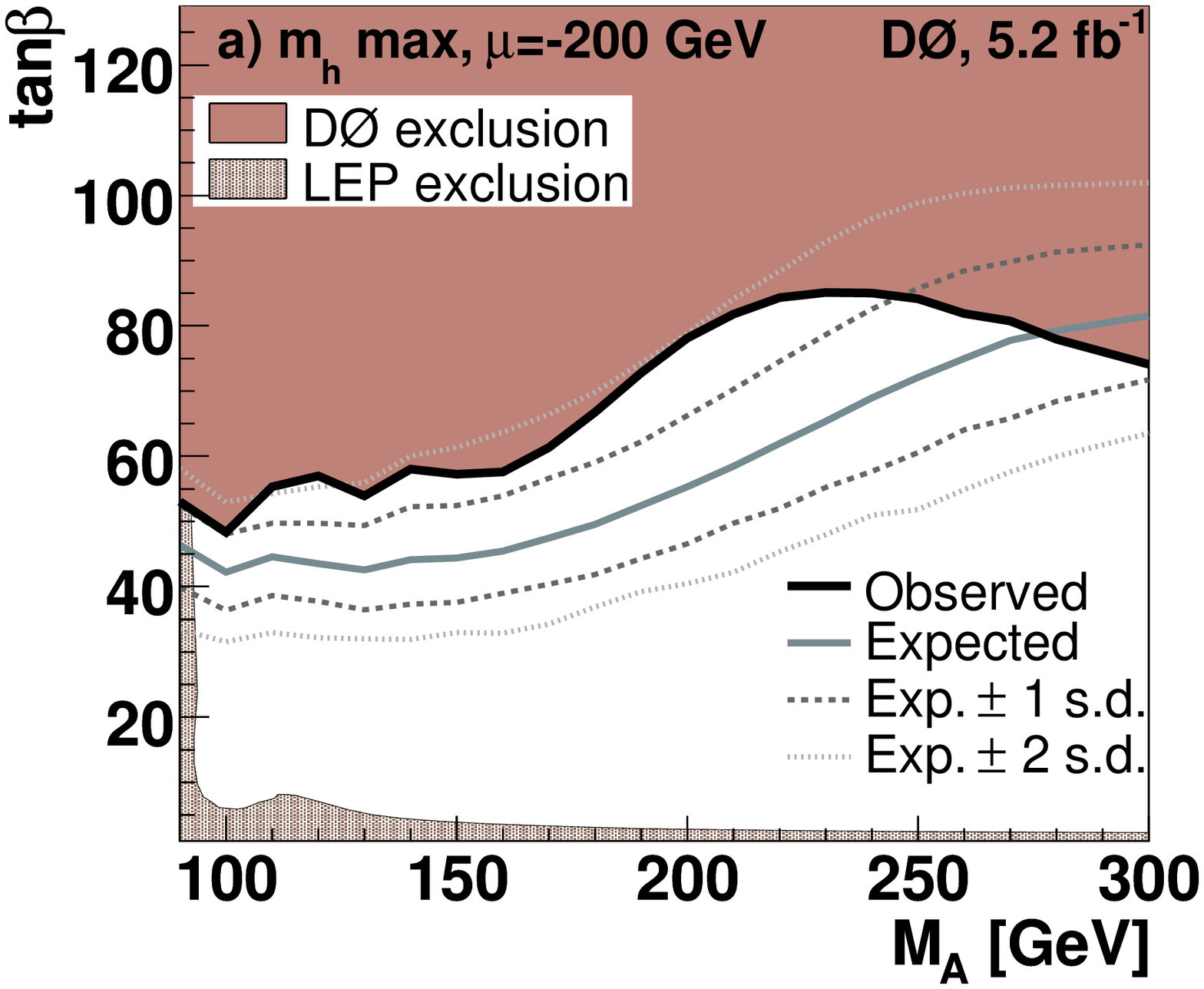}
\includegraphics[width=0.45\linewidth]{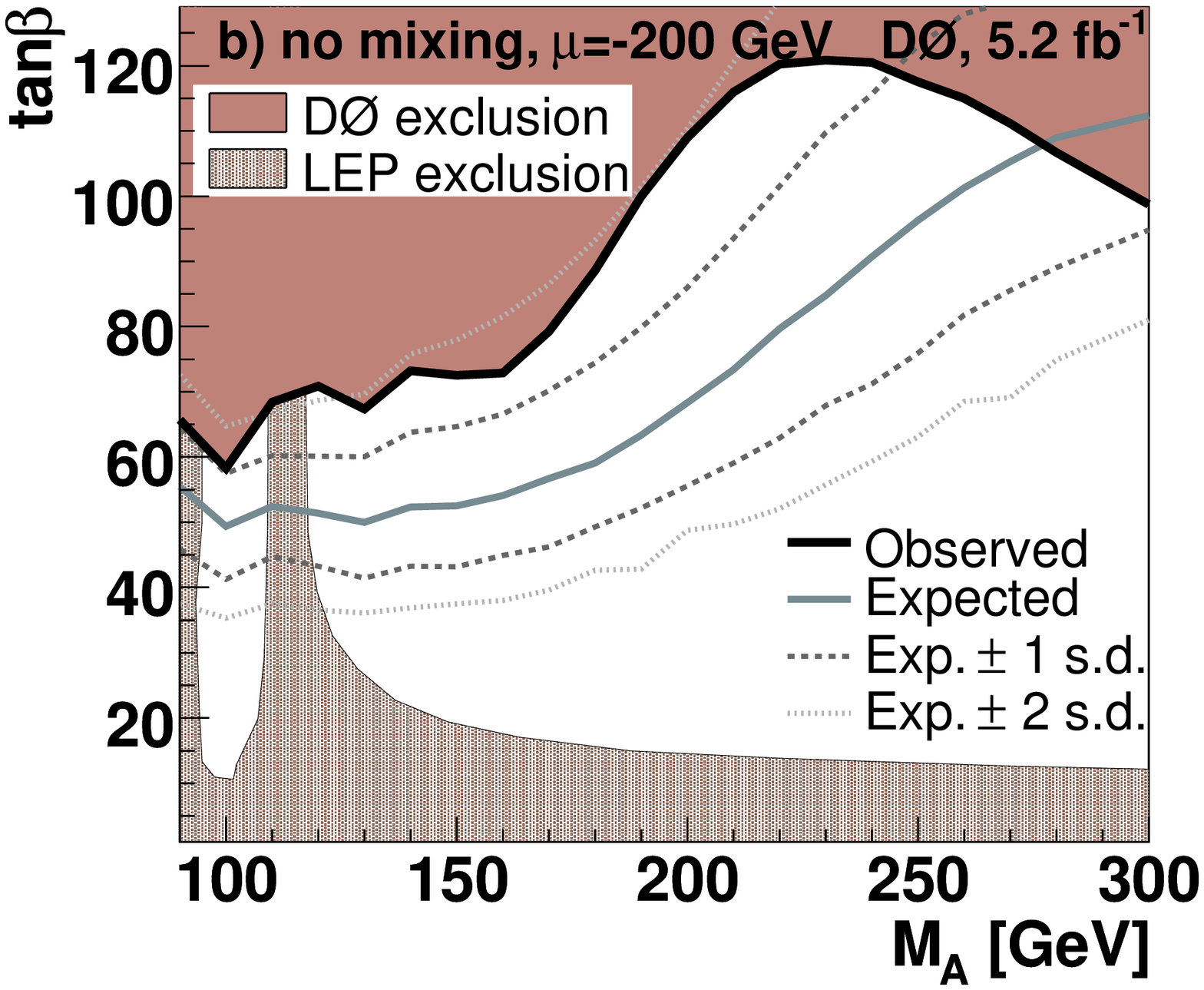}
   \caption{(color online) a) 95\% C.L. lower limit in the $(M_A,\tan\beta)$ plane obtained for the $m^{\rm max}_{h}$, $\mu =-200$~GeV scenario, b) the lower limit for the no-mixing, $\mu =-200$~GeV scenario. The one and two standard deviation bands around the expected limit and the exclusion limit obtained from the LEP experiments are also shown \cite{cite:LEP_exclu}.}
  \label{fig:mssm_nomix_all}
   \ec
\end{figure*}

As a consequence of the enhanced couplings to $b$-quarks at large $\tan\beta$, the total width of the Higgs boson mass also increases with $\tan\beta$. This can have an impact on the search if the width is comparable to or larger than the experimental resolution. To take this effect into account, the width of the Higgs boson is calculated with {\sc feynhiggs}~\cite{feynhiggs} and included in the simulation as a function of the mass and $\tan\beta$ by convoluting a relativistic Breit-Wigner function with the NLO cross section.
The masses and couplings of the Higgs bosons in the MSSM depend, in addition to $\tan\beta$ and $M_A$, on the SUSY parameters through radiative corrections.  Limits on $\tan\beta$ as a function of $M_A$ are derived for two particular scenarios assuming a CP-conserving Higgs sector~\cite{Carena:2005ek}: the $m^{\rm max}_{h}$~\cite{mhmax} and no-mixing~\cite{no-mix} scenarios with a negative or positive value of the Higgs sector bilinear coupling, $\mu$. Figure~\ref{fig:mssm_nomix_all} shows the result interpreted for these two scenarios in the case of $\mu=-200$ GeV. Weaker limits are obtained for the $\mu > 0$ scenarios, due to the decrease in the product of cross section and branching ratio for positive values of $\mu$~\cite{Carena:2005ek}.

The results exclude substantial areas in the MSSM parameter space up to Higgs boson masses of $300$ GeV, under the assumption that a perturbative treatment is valid over the entire region. These are the most stringent limits to date for this topology over this mass range at a hadron collider.

\input acknowledgement.tex
\end{document}

%% file: author_list.tex
\affiliation{Universidad de Buenos Aires, Buenos Aires, Argentina}
\affiliation{LAFEX, Centro Brasileiro de Pesquisas F{\'\i}sicas, Rio de Janeiro, Brazil}
\affiliation{Universidade do Estado do Rio de Janeiro, Rio de Janeiro, Brazil}
\affiliation{Universidade Federal do ABC, Santo Andr\'e, Brazil}
\affiliation{Instituto de F\'{\i}sica Te\'orica, Universidade Estadual Paulista, S\~ao Paulo, Brazil}
\affiliation{Simon Fraser University, Vancouver, British Columbia, and York University, Toronto, Ontario, Canada}
\affiliation{University of Science and Technology of China, Hefei, People's Republic of China}
\affiliation{Universidad de los Andes, Bogot\'{a}, Colombia}
\affiliation{Charles University, Faculty of Mathematics and Physics, Center for Particle Physics, Prague, Czech Republic}
\affiliation{Czech Technical University in Prague, Prague, Czech Republic}
\affiliation{Center for Particle Physics, Institute of Physics, Academy of Sciences of the Czech Republic, Prague, Czech Republic}
\affiliation{Universidad San Francisco de Quito, Quito, Ecuador}
\affiliation{LPC, Universit\'e Blaise Pascal, CNRS/IN2P3, Clermont, France}
\affiliation{LPSC, Universit\'e Joseph Fourier Grenoble 1, CNRS/IN2P3, Institut National Polytechnique de Grenoble, Grenoble, France}
\affiliation{CPPM, Aix-Marseille Universit\'e, CNRS/IN2P3, Marseille, France}
\affiliation{LAL, Universit\'e Paris-Sud, CNRS/IN2P3, Orsay, France}
\affiliation{LPNHE, Universit\'es Paris VI and VII, CNRS/IN2P3, Paris, France}
\affiliation{CEA, Irfu, SPP, Saclay, France}
\affiliation{IPHC, Universit\'e de Strasbourg, CNRS/IN2P3, Strasbourg, France}
\affiliation{IPNL, Universit\'e Lyon 1, CNRS/IN2P3, Villeurbanne, France and Universit\'e de Lyon, Lyon, France}
\affiliation{III. Physikalisches Institut A, RWTH Aachen University, Aachen, Germany}
\affiliation{Physikalisches Institut, Universit{\"a}t Freiburg, Freiburg, Germany}
\affiliation{II. Physikalisches Institut, Georg-August-Universit{\"a}t G\"ottingen, G\"ottingen, Germany}
\affiliation{Institut f{\"u}r Physik, Universit{\"a}t Mainz, Mainz, Germany}
\affiliation{Ludwig-Maximilians-Universit{\"a}t M{\"u}nchen, M{\"u}nchen, Germany}
\affiliation{Fachbereich Physik, Bergische  Universit{\"a}t Wuppertal, Wuppertal, Germany}
\affiliation{Panjab University, Chandigarh, India}
\affiliation{Delhi University, Delhi, India}
\affiliation{Tata Institute of Fundamental Research, Mumbai, India}
\affiliation{University College Dublin, Dublin, Ireland}
\affiliation{Korea Detector Laboratory, Korea University, Seoul, Korea}
\affiliation{CINVESTAV, Mexico City, Mexico}
\affiliation{FOM-Institute NIKHEF and University of Amsterdam/NIKHEF, Amsterdam, The Netherlands}
\affiliation{Radboud University Nijmegen/NIKHEF, Nijmegen, The Netherlands}
\affiliation{Joint Institute for Nuclear Research, Dubna, Russia}
\affiliation{Institute for Theoretical and Experimental Physics, Moscow, Russia}
\affiliation{Moscow State University, Moscow, Russia}
\affiliation{Institute for High Energy Physics, Protvino, Russia}
\affiliation{Petersburg Nuclear Physics Institute, St. Petersburg, Russia}
\affiliation{Stockholm University, Stockholm and Uppsala University, Uppsala, Sweden }
\affiliation{Lancaster University, Lancaster LA1 4YB, United Kingdom}
\affiliation{Imperial College London, London SW7 2AZ, United Kingdom}
\affiliation{The University of Manchester, Manchester M13 9PL, United Kingdom}
\affiliation{University of Arizona, Tucson, Arizona 85721, USA}
\affiliation{University of California Riverside, Riverside, California 92521, USA}
\affiliation{Florida State University, Tallahassee, Florida 32306, USA}
\affiliation{Fermi National Accelerator Laboratory, Batavia, Illinois 60510, USA}
\affiliation{University of Illinois at Chicago, Chicago, Illinois 60607, USA}
\affiliation{Northern Illinois University, DeKalb, Illinois 60115, USA}
\affiliation{Northwestern University, Evanston, Illinois 60208, USA}
\affiliation{Indiana University, Bloomington, Indiana 47405, USA}
\affiliation{Purdue University Calumet, Hammond, Indiana 46323, USA}
\affiliation{University of Notre Dame, Notre Dame, Indiana 46556, USA}
\affiliation{Iowa State University, Ames, Iowa 50011, USA}
\affiliation{University of Kansas, Lawrence, Kansas 66045, USA}
\affiliation{Kansas State University, Manhattan, Kansas 66506, USA}
\affiliation{Louisiana Tech University, Ruston, Louisiana 71272, USA}
\affiliation{Boston University, Boston, Massachusetts 02215, USA}
\affiliation{Northeastern University, Boston, Massachusetts 02115, USA}
\affiliation{University of Michigan, Ann Arbor, Michigan 48109, USA}
\affiliation{Michigan State University, East Lansing, Michigan 48824, USA}
\affiliation{University of Mississippi, University, Mississippi 38677, USA}
\affiliation{University of Nebraska, Lincoln, Nebraska 68588, USA}
\affiliation{Rutgers University, Piscataway, New Jersey 08855, USA}
\affiliation{Princeton University, Princeton, New Jersey 08544, USA}
\affiliation{State University of New York, Buffalo, New York 14260, USA}
\affiliation{Columbia University, New York, New York 10027, USA}
\affiliation{University of Rochester, Rochester, New York 14627, USA}
\affiliation{State University of New York, Stony Brook, New York 11794, USA}
\affiliation{Brookhaven National Laboratory, Upton, New York 11973, USA}
\affiliation{Langston University, Langston, Oklahoma 73050, USA}
\affiliation{University of Oklahoma, Norman, Oklahoma 73019, USA}
\affiliation{Oklahoma State University, Stillwater, Oklahoma 74078, USA}
\affiliation{Brown University, Providence, Rhode Island 02912, USA}
\affiliation{University of Texas, Arlington, Texas 76019, USA}
\affiliation{Southern Methodist University, Dallas, Texas 75275, USA}
\affiliation{Rice University, Houston, Texas 77005, USA}
\affiliation{University of Virginia, Charlottesville, Virginia 22901, USA}
\affiliation{University of Washington, Seattle, Washington 98195, USA}
\author{V.M.~Abazov} \affiliation{Joint Institute for Nuclear Research, Dubna, Russia}
\author{B.~Abbott} \affiliation{University of Oklahoma, Norman, Oklahoma 73019, USA}
\author{B.S.~Acharya} \affiliation{Tata Institute of Fundamental Research, Mumbai, India}
\author{M.~Adams} \affiliation{University of Illinois at Chicago, Chicago, Illinois 60607, USA}
\author{T.~Adams} \affiliation{Florida State University, Tallahassee, Florida 32306, USA}
\author{G.D.~Alexeev} \affiliation{Joint Institute for Nuclear Research, Dubna, Russia}
\author{G.~Alkhazov} \affiliation{Petersburg Nuclear Physics Institute, St. Petersburg, Russia}
\author{A.~Alton$^{a}$} \affiliation{University of Michigan, Ann Arbor, Michigan 48109, USA}
\author{G.~Alverson} \affiliation{Northeastern University, Boston, Massachusetts 02115, USA}
\author{G.A.~Alves} \affiliation{LAFEX, Centro Brasileiro de Pesquisas F{\'\i}sicas, Rio de Janeiro, Brazil}
% ...
%
%
\collaboration{The D0 Collaboration\footnote{with visitors from
$^{a}$Augustana College, Sioux Falls, SD, USA,
$^{b}$The University of Liverpool, Liverpool, UK,
$^{c}$SLAC, Menlo Park, CA, USA,
$^{d}$ICREA/IFAE, Barcelona, Spain,
$^{e}$Centro de Investigacion en Computacion - IPN, Mexico City, Mexico,
$^{f}$ECFM, Universidad Autonoma de Sinaloa, Culiac\'an, Mexico,
and 
$^{g}$Universit{\"a}t Bern, Bern, Switzerland.%
}} \noaffiliation
\vskip 0.25cm

%% file: xsec-table2.tex
\begin{table}
\begin{center}
\begin{tabular}{ddd}
\hline \hline
\multicolumn{1}{c}{$M_A$ [GeV]} & \multicolumn{1}{c}{Observed [pb]}& \multicolumn{1}{c}{Expected [pb]} \\ \hline
90  & 89.5 & 73.9 \\ 
100 & 46.0 & 42.5 \\ 
110 & 55.0 & 34.0 \\ 
120 & 42.0 & 22.6 \\ 
130 & 23.1 & 15.0 \\ 
140 & 17.6 & 10.8 \\ 
150 & 12.4 &  8.05 \\ 
160 &  8.52 &  6.38 \\ 
170 &  7.24 &  5.05 \\ 
180 &  6.37 &  4.11 \\ 
190 &  5.82 &  3.51 \\ 
200 &  5.46 &  2.98 \\ 
210 &  4.43 &  2.64 \\ 
220 &  3.65 &  2.23 \\ 
230 &  2.80 &  2.02 \\ 
240 &  2.19 &  1.81 \\ 
250 &  1.80 &  1.55 \\ 
260 &  1.62 &  1.35 \\ 
270 &  1.31 &  1.23 \\ 
280 &  1.16 &  1.10 \\ 
290 &  0.73 &  1.06 \\ 
300 &  0.63 &  0.95 \\ 
\hline \hline
\end{tabular}
\end{center}
\caption{Model independent 95\% C.L. upper limits on the cross section times branching ratio for the combined 5.2 fb$^{-1}$ analysis. 
\label{tab:xsecComb} }
\end{table}

%% file: acknowledgement.tex
%
We thank the staffs at Fermilab and collaborating institutions,
and acknowledge support from the
DOE and NSF (USA);
CEA and CNRS/IN2P3 (France);
FASI, Rosatom and RFBR (Russia);
CNPq, FAPERJ, FAPESP and FUNDUNESP (Brazil);
DAE and DST (India);
Colciencias (Colombia);
CONACyT (Mexico);
KRF and KOSEF (Korea);
CONICET and UBACyT (Argentina);
FOM (The Netherlands);
STFC and the Royal Society (United Kingdom);
MSMT and GACR (Czech Republic);
CRC Program and NSERC (Canada);
BMBF and DFG (Germany);
SFI (Ireland);
The Swedish Research Council (Sweden);
and
CAS and CNSF (China).

%% file: hbb52.bbl
\begin{thebibliography}{99}
\bibitem{footnote1}  All cross sections are quoted with respect to acceptance requirements on the highest-$p_T$ $b$-quark not from the Higgs decay with $p_T > 12$~GeV and $|\eta| < 5.0$.
\bibitem{footnote2} The events with four $b$-tagged jets in the four jet sample are included with the events with three $b$-tagged jets when determining the background composition.
\bibitem{footnote3} The momentum balance is defined as $|{p_{b_1}} - {p_{b_2}}|/|{p_{b_1}} + {p_{b_2}}|$, where $p_{b_i}$ is the magnitude of the momentum three vector of the $i$th $b$-quark jet.
\end{thebibliography}
